\documentclass[final,5p,times,twocolumn]{elsarticle}
\biboptions{sort&compress}

\usepackage{amssymb}
\usepackage{latexsym}
\usepackage{textcomp}
\usepackage{amsthm}
\usepackage{amsmath}
\usepackage{subfigure}
\usepackage{hyperref}
\usepackage{xcolor}

\journal{Optical Materials}

\begin{document}

\begin{frontmatter}

\title  {Luminescence of BaBrI and SrBrI single crystals doped with Eu$^{2+}$}

\author[igc,isu]{A.~A.~Shalaev\corref{cor1}}
\ead{alshal@mail.ru}
\author[igc,isu]{R. Shendrik\corref{cor1}}
\ead{r.shendrik@gmail.com}
\author[igc]{A.~S.~Myasnikova}
\author[igc,irnitu]{A.~Bogdanov}
\author[igc]{A.~Rusakov}
\author[igc]{A.~Vasilkovskyi}

\address[igc]{Vinogradov Institute of geochemistry SB RAS, 1a Favorskogo str., Irkutsk,Russia, 664033}
\address[isu]{Physics department of Irkutsk state university, 20 Gagarina blvd., Irkutsk, Russia, 664003}
\address[irnitu]{Irkutsk National Researcher Technical University, Lermontov str. 83, Irkutsk, Russia, 664074}
\cortext[cor1]{Corresponding author: Vinogradov Institute of geochemistry SB RAS, 1a Favorskogo, Irkutsk,Russia, 664033}
\begin{abstract}
The crystal growth procedure and luminescence properties of pure and Eu$^{2+}$-doped BaBrI and SrBrI crystals are reported. Emission and excitation spectra were recorded under ultraviolet and vacuum ultraviolet excitations. The energy of the first Eu$^{2+}$ 4f-5d transition and SrBrI band gap are obtained. The electronic structure calculations were performed within GW approximation as implemented in the Vienna Ab Initio Simulation Package. The energy between lowest Eu$^{2+}$ 5d state and the bottom of conduction band are found based on luminescence quenching parameters. The vacuum referred binding energy diagram of lanthanide levels was constructed using the chemical shift model.

\end{abstract}

\end{frontmatter}

\section{Introduction}

For many applications that use inorganic scintillators (such as high energy physics, positron emission tomography, and $\gamma$-ray detection), a high light yield is important along with good energy resolution, and fast response time. However, the search for new scintillators has now slowed down. Therefore, the attention again turned to materials that were previously considered to be luminophores. Moreover, it appears that scintillators based on binary systems have a higher light output than the original matrices. The light output for BaBrI:Eu$^{2+}$ crystals was estimated as 90000 photons/MeV \cite{bourret2012}, whereas in BaBr$_{2}$ and BaI$_{2}$ it is only 58000 and 22000 photons/MeV, respectively \cite{yan2014nuclear}. Light output of SrBrI single crystal doped with Eu$^{2+}$ was estimated about 47000 photons/MeV \cite{bourret2012}. However, optical properties of these crystals has not been investigated well. In a previous article we have studied optical properties of BaBrI and BaClI crystals. The exciton creation energy, band gap and the energy of the lowest 4f-5d transition were estimated and VRBE diagram were constructed \cite{shendrik2017optical}. 

In this article, we present study of the optical properties of pure and Eu-doped SrBrI as well as BaBrI crystals. The optical properties of oxygen centers in BaBrI and SrBrI crystals are also reported. The band gap of SrBrI crystal is estimated. Thermal quenching of Eu$^{2+}$ luminescence is studied. The vacuum referred binding energy (VRBE) diagram for SrBrI crystal is constructed based on experimental data and GGA-PBE calculations. 

\section{Methodology}

\subsection{Crystal Growth}

The compounds of alkaline earth halides are hygroscopic. Therefore, much attention is paid to drying raw materials before crystal growth. The thermogravimetric method (TG) and differential scanning calorimetry (DSC) were used to determine the melting points, the level of hydration and the possible dehydration temperature of the charge materials. The analysis of the charge was carried out using a synchronous thermal analyzer STA 449С Jupiter (NETZSCH). The starting materials were the anhydrous compounds of BaBr$_{2}$, BaI$_{2}$, SrBr$_{2}$ and SrI$_{2}$ (purity 99.9\%, Lanhit, LTD). The stoichiometric mixtures of BaBr$_{2}$+BaI$_{2}$ and SrBr$_{2}$+SrI$_{2}$ were employed. The EuBr$_{3}$ was used for doping the mixtures. Eu$^{2+}$ concentration employed is 0.05 mol.\%. The melting points for BaBrI and SrBrI are 783~$^{\circ}$C and 507~$^{\circ}$C, respectively.
 
Quartz ampoules with special design of 20--30~mm in diameter were developed for bromide mixture drying and melt filtration. This procedure significantly improves the optical quality and scintillation characteristics of the crystals. Quartz ampoule consisted of two sections separated by quartz filter.  

The thermal unit with controlled heating was used to dry the batch in the first section of quartz tube under vacuum. The drying process for raw materials was performed taking into account the TG and DSC data. As a next step, the charge was melted after thorough drying.

The molten compounds were filtered through a quartz filter. The filtered melt was collected in the second section of the ampoule. This section was sealed off under vacuum after filtration and cooling of the melt.  The synthesized compound into this quartz ampoule was used to grow a single crystal.
A vertical 20-zone furnace was used to grow BaBrI:Eu and SrBrI:Eu single crystals via the Bridgman method. The temperature gradient was 10--15~$^{\circ}$C/cm. The rate of the crystal growth was 1~mm/hours. The cooling rate of grown crystals was 10~$^{\circ}$C/hours. After that the growth procedure, the crystals were stored in mineral oil to protect from the atmosphere contamination.

Irregularly shaped 1--1.5~mm thick and 10--15~mm diameter crystal pieces were cleaved from the original boules for further studies. The plates were polished in glove box for the spectroscopic measurements.

\subsection{Luminescence measurements}

Photoluminescence (PL) was measured in vacuum cold-finger cryostat. 
The spectra were recorded with a MDR2 and SDL1 (LOMO) grating monochromator, a photomodule Hamamatsu H6780-04 (185-850 nm), and a photon-counter unit. 
The luminescence spectra were corrected for spectral response of detection channel. 
The photoluminescence excitation (PLE) spectra were measured with a grating monochromators MDR2 and 200~W xenon arc lamp for direct 4f-5d 
excitation and vacuum monochromator VM-2 (LOMO) and Hamamatsu deuterium lamp L7292 for measurements in VUV spectral region. 
The PLE spectra were corrected for the varying intensity of exciting light due. 
The X-ray excited luminescence was performed using an X-ray tube operating at 50 kV and 1 mA. Temperature dependences of luminescence were recorded in vacuum cryostat at linear heating. The heating rate was about 10 K/min.

\subsection{Calculation details}

\textit {Ab Initio} calculations of BaBrI and SrBrI doped with $\mathrm{Eu^{2+}}$ were carried out within 
density functional theory (DFT) using VASP (Vienna 	\textit {Ab Initio} Simulation Package) code~\cite{vasp}. 
The 2\texttimes 2\texttimes 1 (48~atoms) supercell was constructed, 
in which one of $\mathrm{Ba^{2+}}$ or $\mathrm{Sr^{2+}}$ ions was replaced by $\mathrm{Eu^{2+}}$.

Integration within the Brillouin zone was performed on a $\mathrm{\Gamma}$-centered grid of 8 irreducible \textit{k} points.
Geometry optimization was performed with fixed cell dimensions. 
The convergence was achieved if the difference in total energy between the two last iterations was less than $\mathrm{10^{-6}~eV}$.

\section{Results and discussion}

\subsection{Exciton emission}
\begin{figure}[t!]
\centering
\includegraphics[width=0.5\textwidth]{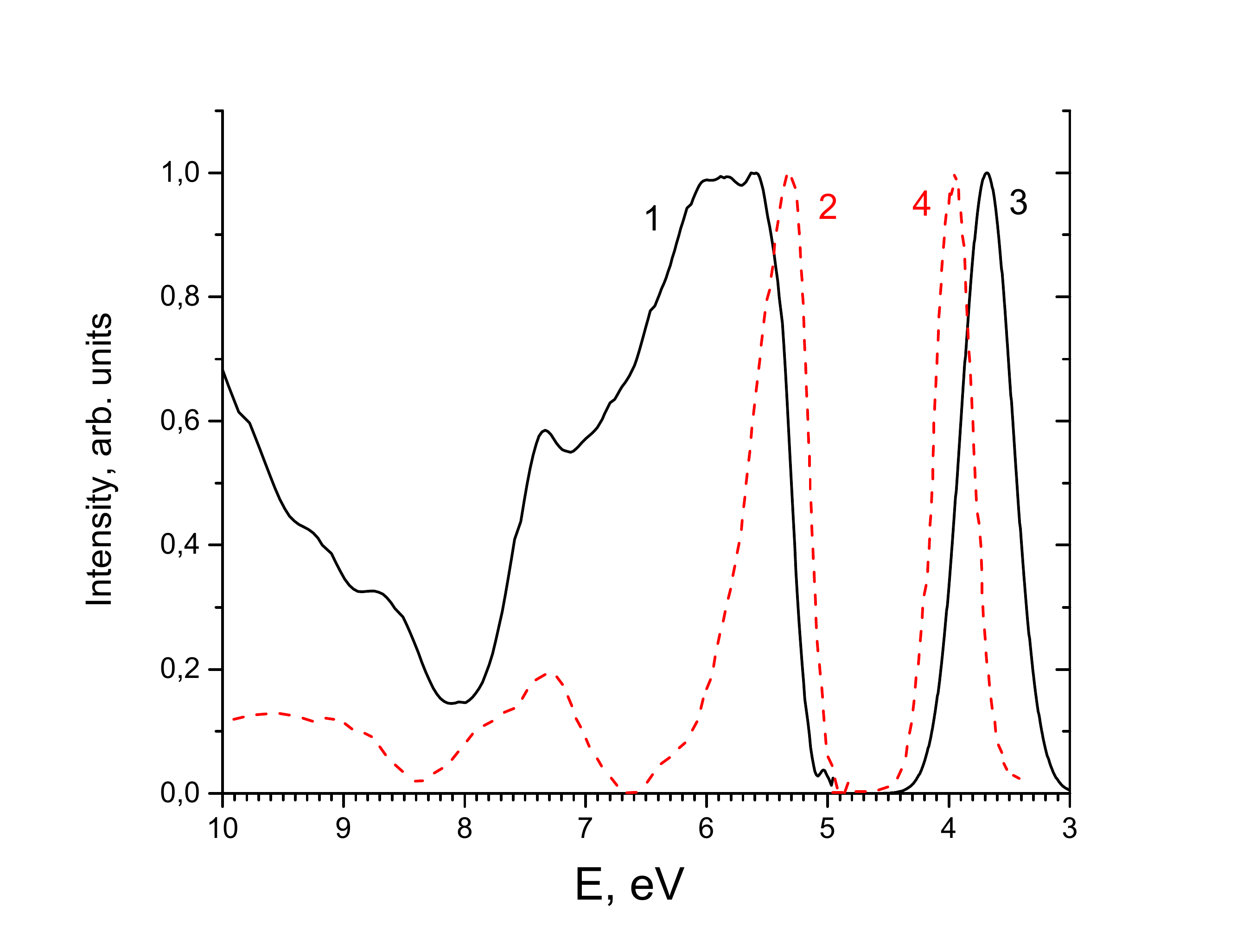}
\caption{Excitation and emission spectra of self-trapped exciton in BaBrI and SrBrI at 78 K. Excitation spectra of SrBrI (solid, curve 1) and BaBrI (dashed, curve 2) crystals monitored at 3.65 eV and 3.95 eV, respectively. Emission spectra were excited at 5.6 eV in SrBrI (curve 3) and 5.3 eV in BaBrI (curve 4) crystals. }
\label{exc-lumi}
\end{figure}

The wide emission band at about 3.7 eV was observed at 78 K in undoped SrBrI crystals when excitation was performed into 5--10~eV region. The excitaton peak (Fig.\ref{exc-lumi}, curve 1) was located at about 5.6 eV. At 5.7 eV a small dip was monitored in excitation spectrum. Intensity of this band decreased in the doped crystals.

In BaBrI crystals the wide band luminescence peaked at 3.9 eV was observed. The excitation peak is located at about 5.3 eV. The wide peaks were found in excitation spectra in 5.5--10 eV range. This luminescence were attributed to self-trapped exciton luminescence \cite{shendrik2017optical}. The same luminescence were observed in the crystals under x-ray excitation.

In SrBrI the most efficient photoluminescence is excited within interband absorption, but excitation peak is located in a higher energy in comparison with BaBrI whereas emission band is shifted to lower energy. This behavior is typical for self-trapped exciton (STE). Similar to other alkali-earth halides \cite{beaumont1970investigation, song2013self, radzhabov1994exciton}, the STE in SrBrI can consist of molecular ion similar to H-center (hole on interstitial bromine or iodine ion) and an F-center-like part (electron trapped by iodine or bromine vacancy). Due to presence of different non-equivalent positions it is possible to observe several types of STE. However, in the present article no attention was paid to this question.

Early we estimated band gap of BaBrI and SrBrI crystals using position of exciton peak. Energy of exciton peak in SrBrI is about 5.7~eV Assuming binding energies of excitons in SrBrI and BaBrI are close, we can estimate band gap of SrBrI crystal following the procedure proposed in \cite{shendrik2017optical}. Exciton creation energy ($E_x$) in SrBrI is obtained at about 5.7~eV. Therefore, band gap is estimated about 6.0~eV.

\subsection{Eu$^{2+}$-luminescence}

\begin{figure}[t!]
\centering
\includegraphics[width=0.5\textwidth]{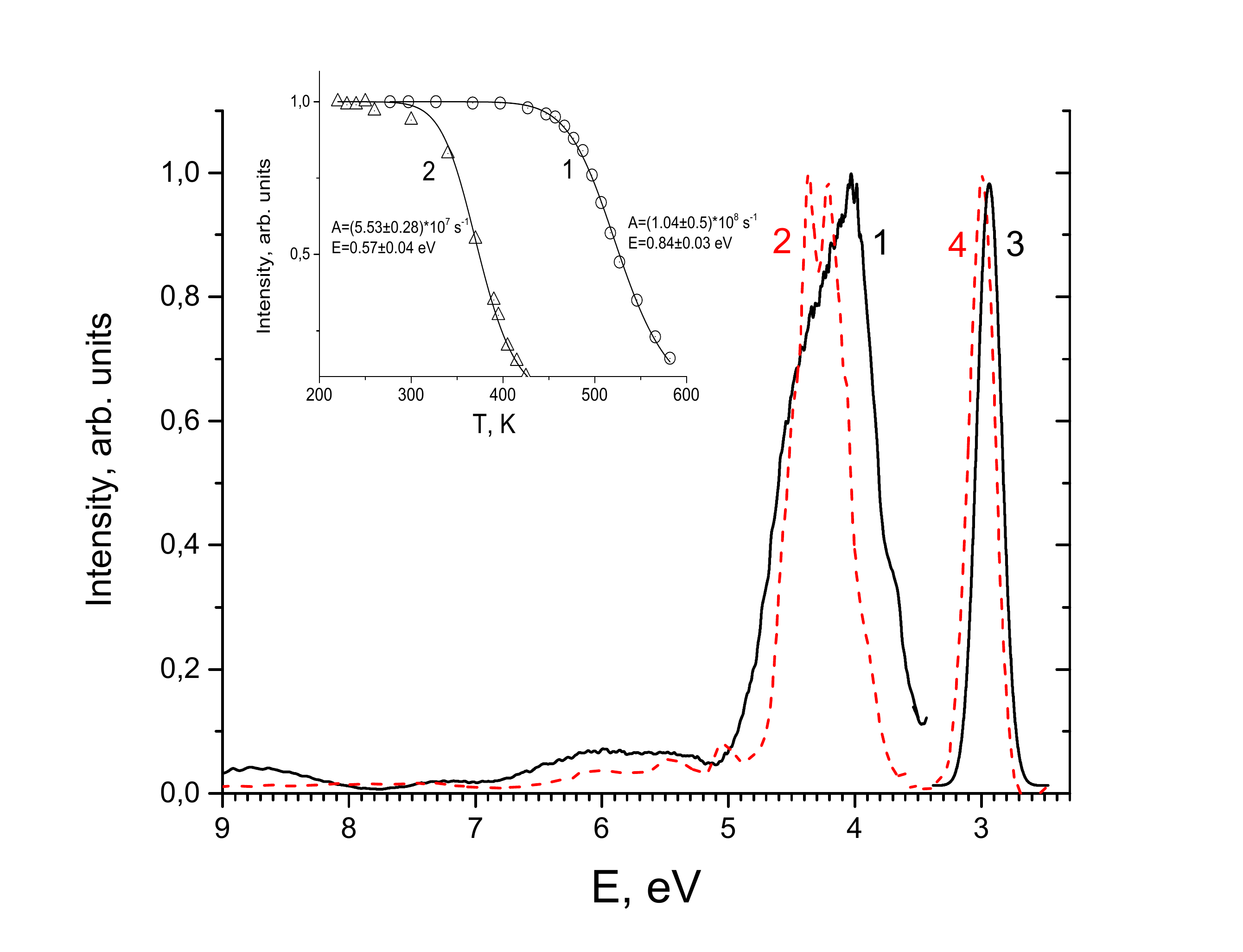}
\caption{Excitation and emission spectra of Eu$^{2+}$ doped BaBrI and SrBrI at 78 K. Excitation spectra of SrBrI (solid, curve 1) and BaBrI (dashed, curve 2) crystals monitored at 3.0 eV and 2.9 eV, respectively. Emission spectra were excited at 4.0~eV in SrBrI (curve 3) and 4.2~eV in BaBrI (curve 4) crystals. In the inset temperature dependences of 5d-4f luminescence intensity of Eu$^{2+}$ ions under 4.0 eV excitation in SrBrI (curve 1) and BaBrI (curve 2) crystals are given.}
\label{eu-lumi}
\end{figure}

The Eu doping leads to appearance of intense luminescence band peaked at about 2.94~eV. This emission spectrum results from 5d-4f transitions in Eu$^{2+}$ ions. The emission peak is slighthly shifted to the lower energy region in comparison with Eu$^{2+}$ emission band in BaBrI crystals at 3.0 eV. Excitation spectra of the emission of Eu-doped BaBrI and SrBrI crystals are given in Fig.~\ref{eu-lumi} The intense excitation peak at 4 eV is attributed to 4f-5d transition in Eu$^{2+}$ ions in SrBrI crystals. The energy of 4f-5d transition in SrBrI is slightly lower than in BaBrI crystals. Using method proposed by Dorenbos \cite{dorenbos2003energy} the lowest energy of 4f-5d transition in BaBrI was estimated as 3.29 eV \cite{shendrik2017optical}. The excitation peak of 5d-4f luminescence in SrBrI shifts by 0.2 eV to the lower energies with respect to BaBrI. Therefore, the lowest energy of 4f-5d transition can be estimated about 3.1 eV. 

Besides the intense 4f-5d excitation band we found that 5d-4f Eu$^{2+}$ emission was excited in exciton and interband transitions region (see fig.~\ref{eu-lumi}). This provides the evidence of energy transfer from the excitons and primary energy carriers (electrons and holes) to the Eu$^{2+}$ ions.
Eu$^{2+}$ emission is more intensive in SrBrI than in BaBrI under exciton excitation. It is due to more overlapping of 4f-5d absorption band and exciton emission in SrBrI than in BaBrI crystals.

The intensity of Eu$^{2+}$ luminescence under intra center direct 4f-5d excitation decreases with increasing temperature. The temperature dependences of Eu$^{2+}$ luminescence for SrBrI and BaBrI crystals are given in the inset in fig.~\ref{eu-lumi}.
In BaBrI crystals Eu$^{2+}$ emission is slightly quenched already at room temperature and intensity dropped to 50\% of the low temperature value at 380~K (T$_{0.5}$=380~K). In SrBrI crystals the Eu$^{2+}$ emission is quenched at higher tmeperatures and T$_{0.5}$ value is about 520 K. The luminescence is quenched following Mott’s law: 
\begin{equation}
\label{Mott_law}
I(T)=\frac{1}{1+w_{0}exp(-E_{A}/k_{B}T)},
\end{equation} 
where $w_0$ is  the  rate  constant  for  the thermally  activated  escape and $E_A$ is the  activation  energy connected with this process. The energies are 0.84$\pm$0.03 eV and 0.5$\pm$0.04 eV for SrBrI and BaBrI, respectively.
 
The different mechanisms of Eu$^{2+}$ luminescence quenching exist. One of the first models proposed by Blase pointed that the quenching of 5d-4f emission in Eu$^{2+}$ could be attributed to a large displacement between the ground and excited states of Eu$^{2+}$ in the configuration coordinate diagram \cite{blasse2012luminescent}. In the other model  Davolos \cite{davolos1989luminescence} assumed that the thermal activation from the 4f$^6$5d$^1$ state to conduction band state causes the quenching in Ba thiogallates. Based on an analysis of the temperature quenching of Eu$^{2+}$ emission in numerous of compounds, Dorenbos showed that the quenching is not due to a large displacement between the ground and excited states of Eu$^{2+}$ in the configuration coordinate diagram. It was proposed that thermal excitation of the 5d electron to conduction band states appears the genuine mechanism. The 5d electron remains bonded in an Eu$^{3+}$ trapped exciton state from which it returns non-radiative to the Eu$^{2+}$ ground state \cite{dorenbos2005thermal}.

\subsection{Luminescence related to oxygen centers}
\begin{figure}[t!]
\centering
\includegraphics[width=0.5\textwidth]{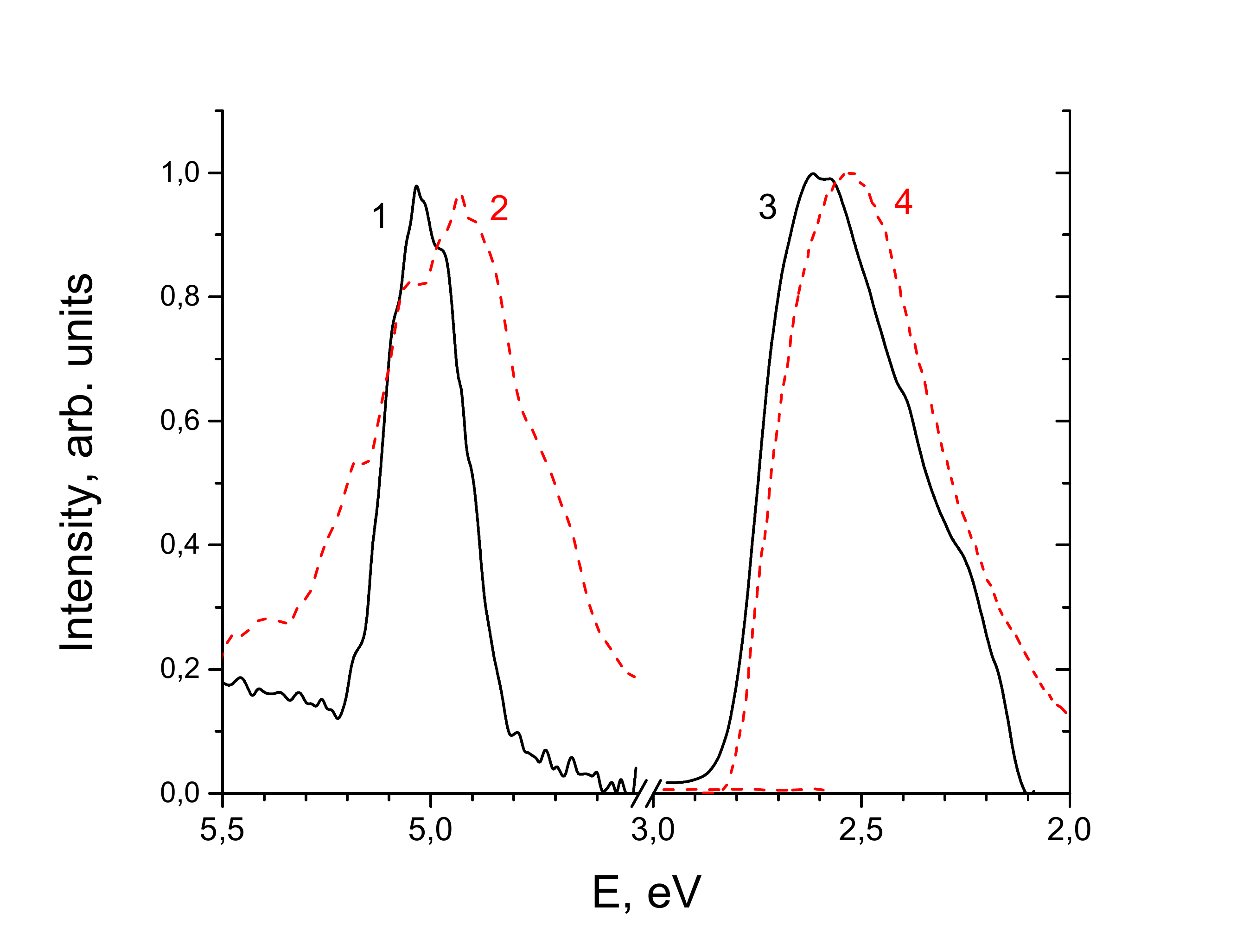}
\caption{. Excitation and emission spectra of oxygen centers in nominally undoped SrBrI and BaBrI crystals. Excitation spectra of SrBrI (solid, curve 1) and BaBrI (dashed, curve 2) crystals monitored at 2.6 eV and 2.5 eV, respectively. Emission spectra were excited at 5.0 eV in SrBrI (curve 3) and 4.9 eV in BaBrI (curve 4) crystals.}
\label{oxy-lumi}
\end{figure}

In some nominally undoped samples of SrBrI and BaBrI crystals the wide band emission peaked at about 2.5 eV was observed. The position of emission and excitation bands slightly depends on the cation type. In SrBrI crystal the wide emission band peaked at 2.6 eV was excited in the band centered at 5.0 eV. In BaBrI crystals emission band centered at 2.5 eV. In the excitation spectrum monitored at 2.5 eV the band peaked at 4.9 eV was found (fig.~\ref{oxy-lumi}). 

The intensity of this emission band depends on preparation of charge for crystal growth. An intensity of this luminescence band was higher in the samples grown by the fast dried and unfiltered raw, where insufficient dehydration took place. When $H_{2}O$ molecules were removed completely from the raw after melt filtration and long 24-hours drying procedures, the intensity of the 2.5 eV emission band is quite low.

This emission bands in BaClI, BaBrI and SrBrI crystals are located in the same spectral region and can be attributed to the oxygen centers \cite{shendrik2017optical}. In strontium iodide a significant part of the oxygen contamination comes from the surface hydrate reactions \cite{pustovarov2012luminescence}. Therefore, concentration of oxygen centers in crystals grown from insufficiently dried substrate should be higher. 

Several types of oxygen centers were found in alkali halide and alkali-earth crystals. In alkali-earth fluorides (CaF$_2$, SrF$_2$, BaF$_2$) and alkali earth dihalides BaFCl, BaFBr oxygen centers (O$^{2-}_{X}$--v$^{\cdot}_{X}$, where X=F, Cl, Br) are formed \cite{egranov1992spectroscopy,radzhabov1994time,radzhabov1995optical,radzhabov1997,pologrudov2008transfer}. Oxygen ion replaces fluorine ion. Charge compensation is performed via anion vacancy formation. The O$^{2-}_{X}$--v$^{\cdot}_{X}$ centers have several absorption bands and one or several broad luminescence bands. It is established that the lowest-energy excitation (absorption) band is approximately constant in crystals with the same alkali metal and follows the Molvo-Ivy dependence in crystals with the same halogen \cite{radzhabov1997,hennl1978optical}.

In the most of alkali-halide crystals, molecular ions of oxygen (O$^{2-}$) were found. The oxygen ions have several absorption bands at about 4-6 eV and a number of well-resolved emission lines at about 1-2 eV \cite{kats1972effect}. The strong ESR signal should be observed in crystals containing oxygen in molecular ion form. The entry form of oxygen is host material related. For example, molecular oxygen ions are practically not observed in alkaline-earth crystals, whereas centers containing oxygen ion and anion vacancy are not observed in alkali-halide crystals \cite{egranov1992spectroscopy,radzhabov1997}.

Oxygen centers related luminescence was observed  in BaBrF and BaClF crystals. The oxygen ion (O$^{2-}$) can occupy Br$^-$ or F$^-$ position in these crystals. Therefore, several types of the centers could be found, there were O$^{2-}_{X}$-v$^{\cdot}_{F}$; O$^{2-}_{X}$-v$^{\cdot}_{X}$ (type II centers); O$^{2-}_{F}$-v$^{\cdot}_{X}$ (type I centers); O$^{2-}_{F}$-v$^{\cdot}_F$ (type III centers), where X=Br, Cl. Here v$^{\cdot}_x$ is vacancy of the anion X. Type I centers, where oxygen substitutes for fluorine ion and anion vacancy occupies bromine or chlorine site, have low energy absorption band at 5.0~eV and wide luminescence band peaked at 2.5 eV \cite{radzhabov1995optical}. Type II centers, where oxygen ion and anion vacancy are located in bromine sublattice, demonstrate absorption band at about 4.2 eV and luminescence at about 2 eV \cite{radzhabov1994time}. Type III centers were not clearly separated in optical absorption spectra and detected only in EPR \cite{eachus1995oxygen}. Ground state of type I centers is mostly formed by 2p orbitals of oxygen and is located in the gap between fluorine and bromine subbands that formed valence band.
The observed luminescence peaked at 2.5 eV and excitation band are very similar to luminescence of type I centers in BaFBr crystals. Intensity of the luminescence depends on dehydration process. Therefore, we can conclude that the 2.5 eV luminescence is related to oxygen vacancy centers. Following to the BaFBr crystals we can expect, that ground state consists from the 2p orbitals of oxygen ions substituting iodine ions but first excited state can be attributed to 1s state of anion vacancy occupied bromine site. The energy of first excitation band is close to band gap. Therefore, if the ground state of oxygen center was in band gap the luminescence would be quenched at room temperature. However, this does not happen in our crystals. Therefore, the ground state of centers containing oxygen ion and anion vacancy should be placed into the gap between iodine and bromine subbands.

\subsection{Calculation results and vacuum referred binding energy (VRBE) diagram of the lanthanide ions in SrBrI}

The ground state of crystals was calculated using spin-polarized generalized gradient approximation (GGA) with the exchange-correlation potential PBE \cite{pbe}. The electronic structure parameters of GGA-PBE calculations are often underestimated, which turned out to be in our case. The calculated band-gap values of the crystals are presented in Table~\ref{all-data} together with experimental data. 

For the calculation of doped crystals, we replaced one of Ba2+ ions in the supercell by the europium ion. To estimate the gap between top of valence band and ground 4f state of Eu$^{2+}$ ion ($E_{CT}$), we applied the method described in \cite{chaudhly} using the Dudarev's approximation PBE+U \cite{Dudarev} for adjusting deposition of 4f levels. In accordance with our previous calculations \cite{shendrik2017optical}, we found that the adequate value of U$_{eff}$ is 2.5. The results of PBE+U calculations are presented in Table~\ref{all-data}. The gap between lowest 5d state of Eu$^{2+}$ and bottom of conduction band ($E_{dC}$) were performed using following relation: 
\begin{equation}
\label{PBE}
E_{dC}= E_{g}-E_{CT}-E_{fd},
\end{equation} 
where $E_{dC}$ -- energy gap between the lowest 4f$^6$5d$^1$ state and bottom of conduction band, $E_{CT}$ -- calculated energy of Eu$^{3+}$ charge transfer band, and $E_{fd}$ -- experimental energy of the first f-d transition in Eu$^{2+}$ ions.

Despite we used ab initio methodology, our calculations are to some extent empirical. So we reexamine the calculation using the GW0 \cite{shishkin} and BSE (Bethe-Salpeter equation)\cite{rohlfing1998electron} methods. The values of the band gap calculated using the GW0 method gives good agreement with the experiment. From the spectra of the imaginary part of dielectric function calculated by BSE method, we estimated the 4f-5d transition energies. All calculated data are presented in Table~\ref{all-data}. Both methods of calculation give $E_{dC}$ energies of about 0.8~eV, which indicates the promise of SrBrI crystals as scintillators. The calculated $E_{dC}$ energies are in good agreement with the experimental ones estimated in the inset of fig.~\ref{eu-lumi}. The calculated values of band gap energies agree well with the experimental data (Table~\ref{all-data}). Therefore, the performed calculation could be used for estimation of band gap and position of 4f levels in other new alkali-earth halides.

\begin{table*}
\begin{center}
 \caption{Calculated band gaps and relative 4f and 5d levels for $\mathrm{Eu^{2+}}$-doped BaBrI and SrBrI crystals. Energies are given in eV.}
\label{all-data}
  \begin{tabular}{|l|c|c|c|c|c|c|}
  \hline
Crystal & \multicolumn{2}{|c|}{Band gap} &  &  &   & \\
& $\mathrm{GW_{0}}$ & Exp. & E$_{CT}$ & E$_{fd}$ & E$_{dC}$ calc. & E$_A$ exp.\\
 \hline
BaBrI &  5.34 & 5.58 & 1.4 & 3.29 & $\sim 0.65$ & 0.57$\pm$0.04 \\
\hline
SrBrI &  5.5 & 6.00 & 1.62 & 3.1 & $\sim 0.85$ & 0.84$\pm$0.03 \\
\hline
  \end{tabular}
\end{center}
\end{table*}

Using measured energies of band gap, exciton creation and 4f-5d transition we can construct vacuum referred binding energy diagram. The diagram shows the binding energies of an electron in the divalent and trivalent lanthanide ion ground and excited states using Dorenbos chemical shift model  \cite{dorenbos2013review}. We have no spectroscopic data on trivalent lanthanides in investigated crystals, therefore we use an Coulomb repulsion energy $U(6,A)$ which defines the binding energy difference between an electron in the ground state of Eu$^{2+}$ with that in the ground state of Eu$^{3+}$ \cite{dorenbos2012modeling, dorenbos2013lanthanide} to define electron binding energy of trivalent lanthanides:
\begin{equation}
\label{dor-f}
U(6,A)=E_{4f}(7,2+,A)-E_{4f}(6,3+,A).
\end{equation}

The ground levels of the 5d configuration of trivalent lanthanides were estimated from the relation within the redshift models \cite{dorenbos2012modeling, dorenbos2003energy}:
\begin{equation}
\label{dor-d}
D(3+,A)=1.563 \cdot D(2+,A)+0.364,
\end{equation} 
where $D(3+,A)$ is redshift of trivalent lanthanide and $D(2+,A)$ is redshift of divalent lanthanide. 
Similar to other alkali-earth iodide and bromide compounds like LaBr$_3$ $U(6,A)$ is estimated about 6.3 eV. The positions of 4f and 5d levels of trivalent lanthanides could be estimated using Eq.\eqref{dor-f}, \eqref{dor-d}.

Dorenbos showed \cite{dorenbos2005thermal} that the energy barrier of thermal quenching ($E_A$ from Eq.\eqref{Mott_law}) of 5d-4f emission is related to energy gap between the lowest 4f$^6$5d$^1$ Eu$^{2+}$ state and the bottom of conduction band. In SrBrI and BaBrI the energy barriers for the luminescence quenching ($E_A$) agree well with the calculated values ($E_{dC}$, see Eq.~\eqref{PBE}) as shown in the Table~\ref{all-data}. Calculated energy E$_{CT}$ for Eu$^{2+}$ ion in BaBrI and SrBrI has lower energy than 4f-4f emission of Eu$^{3+}$ ions which means the absence of charge transfer transition in Eu$^{3+}$ excitation spectrum in these crystals. Therefore, the calculated charge transfer energy $E_{CT}$ and experimental values $E_A$ can be used to estimate the energy gap between the lowest 4f$^6$5d$^1$ state and bottom of conduction band and position of top of valence band related to Eu$^{2+}$ 4f ground state. 

\begin{figure}[t!]
\centering
\includegraphics[width=0.5\textwidth]{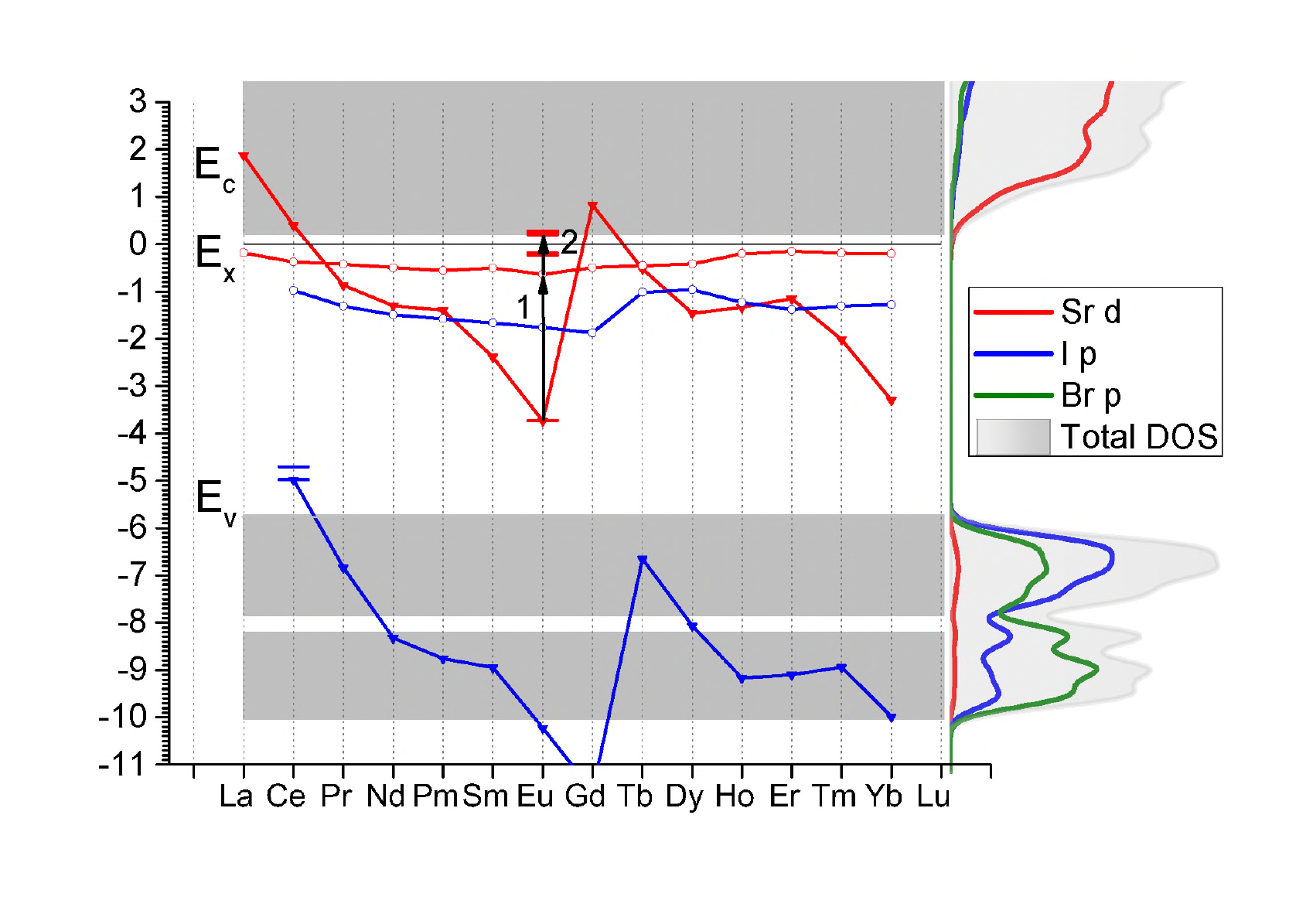}
\caption{The diagram shows the vacuum referred binding energies (VRBE) for electrons of divalent (red) and trivalent (blue) lanthanide ions 
with respect to the vacuum level in SrBrI. The zigzag red curve with triangles connects the 4f$^{n}$ ground state energies 
of the divalent lanthanide ions. The red curve with circles is attributed to their lowest 4f$^{n-1}$5d$^1$ states. 
Blue zigzag curve with triangles respects to 4f$^n$ ground state energies of trivalent lanthanide ions and blue curve 
with circles are their 4f$^{n-1}$5d$^1$ states. Arrow 1 shows the energy of 4f-5d transition and arrow 2 shows the gap between the lowest 4f$^6$5d$^1$ state and bottom of conduction band estimated using experimental $E_{A}$ (Eq.\eqref{Mott_law}) and calculated $E_{dC}$ (Eq.\eqref{PBE}) energies. 
The right inset provides the calculated density of states (DOS) for SrBrI crystal.}
\label{dor-diag}
\end{figure}

The VRBE diagram for SrBrI crystals are given in Fig.~\ref{dor-diag}. The density of states is calculated. It is found the small gap between bromine and iodine subzone similar to BaFBr, BaFCl crystals \cite{radzhabov1995optical, eachus1995oxygen}. Following the diagram we can expect the bright luminescence 5d-4f from Sm$^{2+}$ and Yb$^{2+}$ ions in SrBrI crystals. It would be promising for creation the fast scintillators \cite{alekhin2015luminescence,suta2017decay}. The Ce-doping is also prospecting way due to 5d state of Ce$^{3+}$ ion could be located in sufficiently far from the bottom of conduction band to avoid quenching. 
We estimate that the most ground 4f states of trivalent lanthanide ions except Ce$^{3+}$  lie deep in the valence band. Therefore, we cannot expect to find 5d-4f luminescence of these lanthanides.

\section{Conclusion}
The SrBrI and BaBrI single crystal undoped and doped with 0.05 mol.\% Eu$^{2+}$ ions were grown. The luminescence properties of SrBrI and BaBrI crystals doped with low concentrations of Eu$^{2+}$ ions have been studied. The results obtained by luminescence spectroscopy revealed the presence of europium ions only in divalent form in the BaBrI and SrBrI crystals. 

The intense bands in excitation spectra of SrBrI-Eu crystals at about 3.85-4.2~eV  result from $\mathrm{4f^{7}(^{8}S_{7/2})} \rightarrow 4f^{6}5d^{1}(t_{2g}) $ transitions. The lowest energies of 4f-5d transition in Eu$^{2+}$ ion obtained from the spectra was 3.1~eV for SrBrI. The narrow intense peak at 5.7~eV in the photoluminescence excitation spectrum of the intrinsic emission of SrBrI crystal was due to the creation of excitons. The band gap of SrBrI crystal were estimated about 6.0 eV.

In the undoped crystals wide band luminescence at about 2.5~eV was observed under 5 eV excitation. It corresponds to the centers containing oxygen ion and anion vacancy.

The band gap was estimated using the $\mathrm{GW_{0}}$ approximations. The calculated band gap energies agree with the experimental data. The distance between the lowest 4f$^6$5d$^1$ level of Eu$^{2+}$ ions and top of the valence band has been calculated and estimated using energy barrier of 5d-4f Eu$^{2+}$ luminescence quenching. The calculated and experimental values of energy gap between the lowest 4f$^6$5d$^1$ state and bottom of conduction band are in good agreement The VRBE diagrams of levels of all divalent and trivalent lanthanides ions in SrBrI crystals were constructed on the acquired experimental and theoretical data.

\section*{Acknowledgments}

Crystal growth were supported by RFBR grants 15-02-06514a. Optical spectroscopy of excitons and oxygen centers in undoped crystals and part of calculations were supported by grant of Russian Science Foundation RSF 17-72-10084. The optical spectroscopy of Eu-doped crystals was supported by governmental assignment in terms of Project IX.125.3 (0350-2016-0024; АААА-А17-117101170035-3). The reported study was performed with the equipment set at the Centres for Collective Use ("Isotope-geochemistry investigations" at A.P. Vinogradov Institute of Geochemistry SB RAS).  Generous allotment of computational time from Computational Center of Novosibirsk State University (\url{www.nusc.nsu.ru}) as well as HPC cluster ”Academician V.M. Matrosov” at Irkutsk Supercomputer Centre of SB RAS and ”Academician V.A. Fock” supercomputer at Irkutsk National Research Technical University.

\bibliography{photochromic}

\end{document}